\documentclass[aps,floatfix,nofootinbib,prd, notitlepage,11pt]{revtex4-1}

\usepackage{graphicx}
\usepackage{epic}
\usepackage{eepic}
\usepackage{latexsym}
\usepackage{amssymb,amsmath}
\usepackage{upgreek}

\newcommand{\eq}[1]{(\ref{#1})}
\newcommand{\be}{\begin{equation}}
\newcommand{\ee}{\end{equation}}
\newcommand{\bea}{\begin{eqnarray}}
\newcommand{\eea}{\end{eqnarray}}

\newcommand{\hs}[1]{\hspace{#1 mm}}

\newcommand{\df}{\dot{\phi}}
\newcommand{\dz}{\dot{\zeta}}
\newcommand{\zc}{\zeta_c}
\newcommand{\np}{{\Upphi}}

\newcommand{\lp}{\left\{}
\newcommand{\rp}{\right\}}

\def\cc{\gamma}

\def\d{\delta}

\def\e{\epsilon}

\def\f{\phi}
\def\fr{\frac}
\def\F{\Phi}
\def\vf{\varphi}

\def\l{\lambda}

\def\m{\mu}

\def\r{\rho}

\def\Th{\Theta}

\def\z{\zeta}

\def\del{\partial}

\let\bm=\bibitem
\def\nn{\nonumber}

\begin{document}


\title{On the breakdown of the curvature perturbation $\zeta$ during reheating} 

\author{Merve Tarman Algan}

\email[]{merve.tarman@boun.edu.tr}

\author{Ali Kaya}

\email[]{ali.kaya@boun.edu.tr}

\author{Emine Seyma Kutluk}

\email[]{seymakutluk@gmail.com}

\affiliation{Bo\~{g}azi\c{c}i University, Department of Physics, 34342, Bebek, \.{I}stanbul, Turkey}

\date{\today}

\begin{abstract}

It is known that in single scalar field inflationary models the standard curvature perturbation $\z$, which is supposedly conserved at superhorizon scales, diverges during reheating at times $\dot{\phi}=0$,  i.e. when the time derivative of the  background inflaton field vanishes. This happens because the comoving gauge $\vf=0$, where $\vf$ denotes the inflaton perturbation, breaks down when $\dot{\phi}=0$. The issue is usually bypassed by averaging out the inflaton oscillations but strictly speaking the evolution of $\z$ is ill posed mathematically. We solve this problem in the free theory by introducing a family of smooth gauges that still eliminates the inflaton fluctuation $\vf$ in the Hamiltonian formalism and gives a well behaved curvature perturbation $\z$, which is now rigorously conserved at superhorizon scales. At the linearized level, this conserved variable can be used to unambiguously propagate the inflationary perturbations from the end of inflation to subsequent epochs. We discuss the implications of our results for the inflationary predictions. 

\end{abstract}

\maketitle

\section{\leftline{Introduction}}

Any successful theory of the early universe must explain the origin of the scale free cosmological perturbations seeding the structure in the universe. As a promising candidate, inflation offers a natural  mechanism that yields a scale free spectrum out of the quantum vacuum fluctuations. The well established cosmological perturbation theory studies both the quantum origin and the classical evolution of these perturbations (for a comprehensive review see e.g. \cite{mfb}). An important feature that greatly simplifies the evolution of the cosmological perturbations is the existence of a variable, the curvature perturbation $\z$, which is conserved at superhorizon scales and thus directly connects the observed fluctuations to the very early times. Indeed, under mild technical assumptions it is possible to show that there is always an adiabatic solution to the field equations which is constant at large wavelengths \cite{admo} (a general proof of the conservation of the curvature perturbation without appealing to  perturbation theory has been given in \cite{ly}). 

In this paper, we consider single scalar field inflationary models where the nearly exponential expansion is followed by the period of reheating during which the inflaton field coherently oscillates about the minimum of its potential. In these models, one repeatedly encounters instants in reheating at which $\r+P=0$, where $\r$ and $P$ are the background energy density $\rho=\fr12\df^2+V$ and the pressure   $P=\fr12\df^2-V$. As is well known, the conservation of the curvature perturbation breaks down exactly when $\r+P=0$ (see e.g. \cite{mfb,admo}). Our aim here is to resolve this issue in the single scalar field models at the linearized level. 

A very convenient reformulation of the cosmological perturbation theory is given in \cite{mal}, where the curvature perturbation $\z$ appears as a specific variable in the metric. After fixing the time reparametrization invariance by imposing the so called comoving gauge $\vf=0$  ($\vf$ denotes the inflaton  fluctuation), the scalar degree of freedom is carried out by $\z$, which is naturally conserved at large wavelengths. Indeed, this conservation law can be attributed to a (global) scale invariance and it is valid even in the full quantum theory \cite{c1,c2} (see also \cite{ekw}). However, the comoving gauge $\vf=0$ breaks down when $\df=0$, i.e. when the time derivative of the background inflaton field vanishes, which unfortunately occurs frequently during reheating (note that for a minimally coupled scalar field $\df=0$ gives $\r+P=0$). This problem manifests itself as a singularity in the $\z$ equation of motion, where one sees that a generic smooth initial data diverges in time as $\df\to0$ and the Bunch-Davies mode function is not an exception. Strictly speaking, the evolution of $\z$ is ill defined during reheating and it fails to be the demanded conserved quantity. 

In studying the possible effects of reheating on the cosmological perturbations, one unavoidably faces the breakdown of $\z$. The standard way of smoothing out the singularities (or as it is sometimes referred, the spikes) of  $\z$ is to carry out some form of averaging over inflaton oscillations. For example, an important problem is to see whether the oscillations of the inflaton background causes a real superhorizon evolution of the cosmological fluctuations and $\z$ is the key variable in answering this question. In that case, $\z$ can be smoothed out by replacing the function $\df^2$ by its time average over the inflaton oscillations $<\hs{-1}\df^2\hs{-1}>$  (see e.g. \cite{rme1,rme2,rme3}). Note that this averaging avoids $\r+P=0$, since the mean pressure generically vanishes during reheating. In a different setting, the same procedure can be used to cure the entropy mode loop corrections  to the scalar power spectrum during reheating \cite{a2}. However, this method is not mathematically well justified due to the substitutions like $1/\df^2\to1/<\hs{-1}\df^2\hs{-1}>$. Of course, it is possible to define other variables which are well behaved during reheating like the gauge invariant Sasaki-Mukhanov variable or the inflaton fluctuation field $\vf$ that carries the scalar mode in the  $\z=0$ gauge.  The main problem in using  these variables is that they are not necessarily conserved at large wavelengths and consequently they involve nontrivial superhorizon evolutions during reheating. As a result, the details of the reheating or other subsequent stages become important in determining the cosmological observations. 

In this work we take the background inflaton oscillations at reheating seriously, and try to find an honestly conserved variable, which is also well defined during reheating. As noted above, and as we will discuss below, the main reason for $\z$ becoming an ill defined variable is the breakdown of the comoving gauge $\vf=0$ at times $\df=0$. Thus, the main problem is to find a smooth gauge condition that still eliminates the inflaton fluctuation $\vf$ from the dynamics. The Hamiltonian formalism is best suited for this analysis and we indeed find a  smooth family of gauge conditions that can be used to remove $\vf$ from the system, leaving a well behaved and conserved curvature perturbation $\z$. Having obtained this smooth variable, we also find transformations that directly relate it to the old singular $\z$ and to the Newtonian gravitational potential in the longitudinal gauge, which determines the power spectrum of the density fluctuations. As we will discuss, our results put some of the standard inflationary predictions on a firmer mathematical basis. 

\section{\leftline{Single Scalar Field Models and Cosmological Perturbations}\\ \leftline{in the  Comoving Gauge}}\label{sec2}

In this section, following \cite{mal} we review the derivation of the equations of cosmological perturbations in the single scalar field models. We consider a generic model with the potential $V(\f)$, where $\f$ denotes the inflaton field that is minimally coupled to gravity. In such a model, the background field equations can be written as
\bea 
&&\ddot{\f}+3H\df+V_\f=0,\nn\\
&&3H^2=\fr12 \df^2+V,\label{be}\\
&&\dot{H}=-\fr12 \df^2,\nn
\eea
where $V_\f=\del V/\del\f$, the background metric is given by 
\be\label{bm}
ds^2=-dt^2+a^2d\vec{x}^2,
\ee
$H$ is the Hubble parameter $H=\dot{a}/a$, dots denote ordinary time derivatives with respect to $t$ and we set $8\pi G=1$. For now, the background evolution is completely arbitrary, i.e. we make no assumptions like the slow-roll approximation, and any set of fields obeying \eq{be} is allowed for the upcoming analysis. Later on we will focus on inflation that is followed by reheating. 

Using the ADM formalism, the action governing the dynamics of this theory can be written as 
\be\label{a}
S=\fr12 \int \sqrt{h} \left[N\left(R^{(3)}-2V-h^{ij}\del_i\phi\del_j\f\right)+\fr{1}{N}\left(K_{ij}K^{ij}-K^2+(\dot{\f}-N^i\del_i\f)^2\right)\right],
\ee
where the metric is parametrized as
\be
ds^2=-N^2dt^2+h_{ij}(dx^i+N^i dt)(dx^j + N^j dt),
\ee
$R^{(3)}$ is the Ricci scalar of $h_{ij}$, $K_{ij}=\fr12(\dot{h}_{ij}-D_i N_j-D_j N_i)$, $K=h^{ij}K_{ij}$ and $D_i$ is the derivative operator of $h_{ij}$. Note that the indices in these expressions are manipulated by the spatial metric $h_{ij}$. For a homogeneous inflaton field $\f(t)$ and the metric \eq{bm}, one may easily obtain \eq{be} from the action \eq{a}.

The cosmological perturbations can be introduced as 
\bea
&&h_{ij}=a^2e^{2\zeta}(e^\cc)_{ij},\hs{5}\cc_{ij}=\cc_{ji},\hs{5}\cc_{ii}=0,\nn\\
&&\f=\f(t)+\vf,\label{cpert}\\
&&N=1+n,\nn\\
&&N^i=0+n^i,\nn
\eea
where $a(t)$ and $\f(t)$ are assumed to obey the background equations \eq{be}. Note that $\cc_{ij}$ is defined to be traceless and the determinant of $h_{ij}$ is parametrized by $\z$. At the linearized level, the infinitesimal coordinate transformations that are generated by the vector field $k^\m=(k,k^i)$ yield the following gauge transformations
\bea
&&\d\z=Hk+\fr13 \del_ik^i,\nn\\
&&\d\vf=\df k,\nn\\
&&\d n=\dot{k},\label{gt}\\
&&\d n^i=\dot{k}^i-\fr{1}{a^2}\del_i k,\nn\\
&&\d\cc_{ij}=\del_ik_j+\del_jk_i-\fr23\d_{ij}(\del_mk^m),\nn
\eea
where the indices are raised and lowered by the Kronecker delta $\d_{ij}$. 

Using \eq{cpert} in \eq{a}, one gets the action for cosmological perturbations. In this paper, we only consider the {\it free theory} and thus it would be enough to determine the quadratic action. The equations for the lapse $n$ and the shift $n^i$ are algebraic and these can be used to determine $n$ and $n^i$ in terms of the other fields. On the other hand, the gauge freedom \eq{gt} can be fixed completely by imposing
\be\label{cgeski}
\vf=0,\hs{5}\del_i\cc_{ij}=0.
\ee
This is called the comoving gauge where the scalar mode is carried out by $\z$ and as usual $\cc_{ij}$ represents the gravitational waves. In that case, the lapse and the shift can be solved at the linear order as \cite{mal} 
\be \label{ls}
n=\fr{\dot{\z}}{H},\hs{5}n^i= \fr12 \del^{-2}\del_i\left[\fr{\df^2}{H^2}\dz^2-\fr{2}{a^2H}\del^2\z\right].
\ee
Note that the solution for the shift $n^i$ involves the Green function for the flat space Laplacian $\del^2$ and thus it is non-local in the position space. Nevertheless, \eq{ls} can easily be understood in momentum space. As discussed in \cite{mal}, to obtain the action up to cubic order in fluctuations it is enough to determine $n$ and $n^i$ at the linear order. One may now use \eq{ls} back in the action \eq{a} to eliminate $n$ and $n^i$ completely, leaving a Lagrangian for the three physical degrees of freedom carried out by $\z$ and $\cc_{ij}$. A straightforward calculation shows that the quadratic actions decouple $S^{(2)}=S_\z^{(2)}+S_\cc^{(2)}$, where
\bea
&&S_\z^{(2)}=\fr12\int a^3\fr{\dot{\f}^2}{H^2}\left[\dot{\z}^2-\fr{1}{a^2}(\del\zeta)^2\right],\label{qac}\\
&&S_\cc^{(2)}=\fr18\int a^3\left[\dot{\cc}_{ij}^2-\fr{1}{a^2}(\del_k\cc_{ij})^2\right].\label{cceq}
\eea
It is important to note that no slow-roll approximation is used in deriving these actions, i.e. they are valid provided that the  background field equations \eq{be} hold. 

The full set of equations must be invariant under the transformation $a\to\l a$ that corresponds to a constant scaling of the spatial coordinates $\vec{x}$. From \eq{cpert}, where the perturbations are introduced,  the scaling $a\to\l a$ can be interpreted as a shift  of $\z$ as $\z\to\z+\ln\l$. This shows that the constant $\z$ configuration, which is generated by this shift from $\z=0$, should solve the $\z$ equation of  motion to all orders in the perturbation theory \cite{mal}. This special property leads to the appreciated superhorizon conservation law. 

\section{\leftline{The Breakdown of $\z$ During Reheating}}

\begin{figure}
\centerline{\includegraphics[width=8cm]{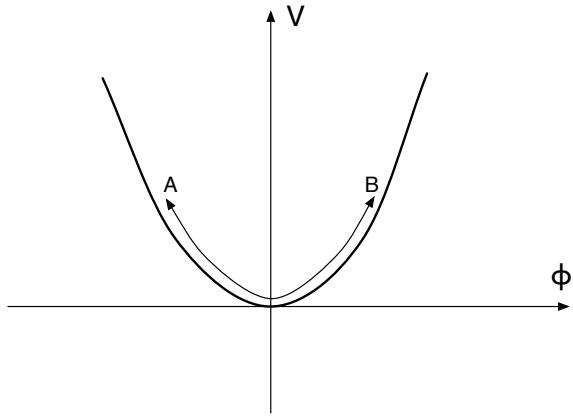}}
\caption{During reheating the inflaton oscillates about the minimum of its potential and at the turning points $A$ and $B$, one has $\df=0$ causing a problem for the $\z$ evolution. In time, the amplitude of oscillations decreases i.e. $A$ and $B$ approaches to the origin.} 
\label{f1}
\end{figure}

One sees from the action \eq{qac} that the evolution of $\z$ becomes singular at $\df=0$. Indeed, the field equation of $\z$ can be found as
\be\label{ze}
\ddot{\z}+\left[3H+\fr{2\ddot{\f}}{\df}-\fr{2\dot{H}}{H}\right]\dot{\z}-\fr{1}{a^2}\del^2\z=0,
\ee
where the second term in the square brackets involving $1/\df$ is problematic: Given an initial generic data set $\z(t_0,x)$ and $\dz(t_0,x)$, the solution developed by \eq{ze} is singular where $\ddot{\z}$ necessarily diverges\footnote{One also sees from the action \eq{qac} that the propagator of $\z$ blows up in quantum theory at $\df=0$.} in time at $\df=0$. For the class of inflationary models we are considering, one frequently encounters instants for which $\df=0$, see Fig. \ref{f1}. 

The singular behavior emerging from \eq{ze} can nicely be illustrated for the superhorizon modes. Introducing the standard Fourier space components $\z_k$, the two solutions of \eq{ze} for negligible wavenumber $k$ can be found as
\be\label{0o}
\z_k\simeq \z_k^{(0)}+c_k\int^t dt'\fr{H^2(t')}{a^3(t')\df^2(t')},
\ee
where $\z_k^{(0)}$ and $c_k$ are two constants, which are fixed by the vacuum chosen and the time of horizon crossing $t_k$. Since $\f$ is oscillating, one has $\ddot{\f}\not=0$ near $\df=0$, and thus the integral in the second solution diverges\footnote{While the integral $\int dx/x$ can be cured by taking the principal value, this method does not work for $\int dx/x^2$.} like $\int d(\df)/\df^2=\int^0 dx/x^2$. One may suggest to discard the second solution since it is decaying during inflation and becomes completely negligible for the modes of cosmological interest (the Wronskian condition, which is required for the canonical commutation relations to hold, implies $\z_k^{(0)}c_k^*-\z_k^{(0)*}c_k=i$, and thus it is not possible to set $c_k=0$). However, this decaying solution naturally appears in the loop effects of entropy modes during reheating \cite{a2}. Moreover, the modes leaving  the horizon just before inflation and reentering the horizon in reheating might have important physical effects (for instance, they may cause primordial black holes to form, see e.g. \cite{bh}) and for those modes the decaying solution is not completely negligible. 

On the other hand, the ``constant" mode is not safe either. The first order derivative correction to the superhorizon mode can be obtained from \eq{ze} as \cite{w1}
\be\label{1o}
\z_k\simeq\z_k^{(0)}\left[1-k^2\int^t dt'\fr{H^2(t')}{a^3(t')\df^2(t')}\int^{t'}dt''a(t'')\fr{\df^2(t'')}{H^2(t'')}\right].
\ee
Among other things, this correction determines the semiclassicality of the superhorizon modes \cite{sc}, and we see that it once more blows up  as the first integral passes through the singularity of the integrand at $\df(t')=0$. 

In general, it is easy to prove that any smooth generic initial data set, which is evolved by \eq{ze}, would yield a solution that diverges like $\z\propto 1/\df$ near $\df=0$. To see this, one may introduce a new variable by $\vf=-(\df/H)\z$. An easy calculation then shows that the field equation of $\vf$ becomes completely regular even at $\df=0$. Therefore, $\z$ must blow up like $1/\df$ as $\df\to0$. Indeed, $\vf$ is nothing but the inflaton fluctuation, which is an entirely well defined variable. Using the gauge freedom \eq{gt}, an arbitrary configuration $(\vf,\z)$ can be transformed to $(0,\z)$ or $(\vf,0)$ by choosing $k=-\vf/\df$ or $k=-\z/H$, respectively. As a result, the gauge transformation from $(0,\z)$ to $(\vf,0)$ gives the relation $\vf=-(\df/H)\z$. These comments also explain the main reason for the breakdown of $\z$. Namely, to impose the comoving gauge $\vf=0$ one must take $k=-\vf/\df$ in \eq{gt}, which gives a diverging gauge parameter at $\df=0$. Therefore, the comoving gauge is {\it not an allowed condition} if $\df=0$. 

\section{\leftline{A Family of Smooth Gauges}}

In this section, we discuss how one can eliminate the inflaton perturbation $\vf$ from the dynamics by imposing smooth gauge fixing conditions. Whether such smooth conditions exist and whether they can be used to remove $\vf$ from the system are not evident in the Lagrangian formulation, where the comoving gauge \eq{cg} looks like the only possible option. Therefore, we switch to the Hamiltonian formalism where the gauge invariance is connected to the existence of constraints in the phase space (see e.g. \cite{prop} for a similar Hamiltonian analysis). This would give a better geometric picture about the dynamical evolution of the system and it may suggest better alternatives to the comoving gauge. 

Since we would like to work out  gauge fixing in the Hamiltonian formalism,  we keep all  11 variables $\z$, $\cc_{ij}$, $\vf$, $n$ and $n^i$ in the Lagrangian. Note that $\cc_{ij}$ is traceless and thus it contains only 5 independent degrees of freedom. Plugging  \eq{cpert} into \eq{a} {\it without imposing any conditions}, the quadratic action of all fluctuations (recall that in this paper we only consider the free theory) can be obtained as
\bea
S^{(2)}=&&\int \fr18 a^3\dot{\cc}_{ij}\dot{\cc}_{ij}-\fr18 a(\del_i\cc_{jk})(\del_i\cc_{jk})+\fr14 a (\del_i\cc_{ik})(\del_j\cc_{jk})-3a^3\dz^2+a\del_i\z\del_i\z+\fr12 a^3\dot{\vf}^2\nn\\
&&-\fr12 a\del_i\vf\del_i\vf-\fr12 a^3\overline{V}_{\f\f}\,\vf^2+\fr12 a\z\del_i\del_j\cc_{ij}-3a^3\df\dz\vf\label{qa}\\
&&-\fr12 a^3\del_in_j\dot{\cc}_{ij}+2a^3\dz\del_in_i+\fr14 a^3(\del_in_j)(\del_in_j)-\fr14 a^3(\del_in_i)^2-a^3\df n_i\del_i\vf\nn\\
&&\fr12 a^3n\left[\fr{1}{a^2}\del_i\del_j\cc_{ij}-\fr{4}{a^2}\del^2\z-2\overline{V}_\f\vf-2\df\dot{\vf}+12H\dz-4H\del_i n_i\right]- a^3\overline{V}n^2,\nn
\eea
where $V_\f=\del V/\del\f$,  all index manipulations are carried out by the Kronecker delta $\d_{ij}$ (like $n_i=n^i$)  and an over-line indicates that the function is evaluated in the background solution. The canonical momenta conjugate to $\cc_{ij}$, $\z$, $\vf$, $n$ and $n^i$, which are respectively denoted by $\Pi_{ij}$, $P_\z$, $P_\vf$, $P_n$ and $P_i$,  can be found from this quadratic action as 
\bea
&&\Pi_{ij}=\fr14 a^3\dot{\cc}_{ij}-\fr14 a^3\left[\del_in_j+\del_jn_i-\fr23\d_{ij}\del_kn_k\right],\nn\\
&&P_\z=-6a^3\,\dz-3a^3\df\,\vf+2a^3\del_in_i+6Ha^3\,n,\nn\\
&&P_\vf=a^3\,\dot{\vf}-a^3\df\, n,\label{conm}\\
&&P_n=0,\nn\\
&&P_i=0.\nn
\eea
One can then apply the standard Legendre transformation to obtain the Hamiltonian\footnote{While $H$ is the Hamiltonian, ${\cal H}$ denotes the respective density.  With some abuse of terminology, we also refer ${\cal H}$ as the Hamiltonian.} of the system as $H=\int d^3x\, {\cal H}$:
\bea
{\cal H}=&&\fr{2}{a^3}\,\Pi_{ij}\Pi_{ij}+\fr{1}{8}\,a\,\del_i\cc_{jk}\del_i\cc_{jk}-\fr{1}{4}\,a\,\del_j\cc_{jk}\del_i\cc_{ik}-a\del_i\z\del_i\z-\fr{1}{12a^3}P_\z^2-\fr{1}{2}\df\,\vf P_\z +\fr{1}{2a^3}P_\vf^2\nn\\
&&+\fr12 a^3\overline{V}_{\f\f}\vf^2-\fr{3}{4}a^3\df^2\,\vf^2+\fr{1}{2}\,a\,\del_i\vf\del_i\vf-\fr{1}{2}\,a\,\z\del_i\del_j\cc_{ij}-n\F-n^i\F_i,\label{h}
\eea
where  
\bea
&&\F_i=2\del_j\Pi_{ji}+\fr13 \del_iP_\z,\label{c1}\\
&&\F=\fr{1}{2}\,a\,\del_i\del_j\cc_{ij}-2\,a\,\del^2\z+a^3\,\ddot{\f}\,\vf-\df P_\vf-HP_\z.\label{c2}
\eea
Since not all components of $\cc_{ij}$ are independent due to trace-freeness, it is convenient to introduce a basis in the space of  $3\times3$, traceless, symmetric matrices, which we denote by $T^a_{ij}$, $a,b=1,2,..,5$. One may normalize these matrices  to satisfy 
\bea
&&T^a_{ij}T^b_{ij}=\d^{ab},\label{t}\\
&&T^a_{ij}T^a_{mn}=\fr12\left[\d_{im}\d_{jn}+\d_{in}\d_{jm}-\fr23\d_{ij}\d_{mn}\right].\nn
\eea
We expand $\cc_{ij}$ and $\Pi_{ij}$ as $\cc_{ij}=\cc_aT^a_{ij}$ and $\Pi_{ij}=\Pi_aT^a_{ij}$, and treat $\cc_a$ and $\Pi_a$ as  independent variables. 

The canonically conjugate variables obey the standard Poisson bracket relations:
\bea
&&\left\{ \z(t,x),P_\z(t,y)  \right\}=\d^3(x-y),\nn\\
&&\left\{\vf(t,x),P_\vf(t,y)\right\}=\d^3(x-y),\\
&&\left\{\cc_{ij}(t,x),\Pi_{mn}(t,y)\right\}=\fr12 \d^3(x-y)\left[\d_{im}\d_{jn}+\d_{in}\d_{jm}-\fr23\d_{ij}\d_{mn}\right],\nn
\eea
where in getting the last relation we have used the properties of the matrices $T^a_{ij}$ given in \eq{t}.  The constancy of the primary constraints in \eq{conm} under the Hamiltonian flow, i.e. $P_n=0$ and $P_{i}=0$, implies 
\be
\F_i=0,\hs{10}\F=0. 
\ee
As usual, $n$ and $n^i$ become non-dynamical Lagrange multipliers. On the other hand, the first class constraints $\F_i$ and $\F$, which are are related to the gauge invariance \eq{gt}, obey 
\be
\lp \F,\F_i\rp=0.
\ee
A direct calculation shows that the constraints are preserved under the Hamiltonian flow: 
\bea
&&\fr{d\F_i}{dt}=\lp \F_i,H\rp=0\nn\\
&&\fr{d\F}{dt}=\lp \F,H\rp+\fr{\del\F}{\del t}=0.
\eea
This is a highly nontrivial consistency check of the expressions given above. One must note that the constraint $\F$ has an explicit time dependence through the background fields $a(t)$ and $\f(t)$, which are assumed to obey the equations of motion \eq{be}. For a given vector field $k^\m=(k,k^i)$ one may define $\Th=\int d^3 x (k\F+k^i\F_i)$, which generates the gauge transformations in \eq{gt}
\bea
&&\d\z=\lp \Th,\z\rp =Hk+\fr13 \del_ik^i,\nn\\
&&\d\vf=\lp \Th,\vf\rp =\df k,\label{gtps}\\
&&\d\cc_{ij}=\lp \Th,\cc_{ij}\rp=\del_ik_j+\del_jk_i-\fr23\d_{ij}(\del_mk^m).\nn
\eea
The transformations of the conjugate momenta can be found as 
\bea
&&\d P_\z=\lp \Th,P_\z\rp =-2a\del^2k,\nn\\
&&\d P_\vf=\lp \Th,P_\vf\rp =a^3\ddot{\f}\,k\label{gtpsm}\\
&&\d \Pi_{ij}=\lp \Th,\Pi_{ij}\rp=\fr12 a\left[\del_i\del_jk-\fr13\d_{ij}\del^2k\right].\nn
\eea
Since we treat $n$ and $n^i$ as non-dynamical Lagrange multipliers, their gauge transformations can be imposed from the invariance of the action $S=\int (p\dot{q}-H)$ under \eq{gtps} and \eq{gtpsm} (see e.g. \cite{t}), which gives the corresponding equations in \eq{gt}, i.e.
\bea
&&\d n=\dot{k},\nn\\
&&\d n^i=\dot{k}^i-\fr{1}{a^2}\del_i k.\label{gtn}
\eea
One may check that the gauge transformations \eq{gtps}, \eq{gtpsm} and \eq{gtn} are consistent with \eq{conm}. 

After obtaining the Hamiltonian and the first class constraints, we now proceed with the gauge fixing. As is well known, a function that has a non-zero Poisson bracket with a first class constraint defines an allowed gauge for the corresponding invariance. After finding such a suitable gauge condition, one may prefer to work either in the {\it full phase space} by replacing the standard Poisson brackets with the Dirac brackets \cite{dirac}, which are defined by the solutions of the Lagrange multipliers, or  rather ``solve the constraints"  to obtain {\it a reduced phase space} and {\it a reduced Hamiltonian} \cite{cs}. The equivalence of these two procedures, when there is no explicit time dependence, has been proved in \cite{cs}. In the appendix \ref{ap1}, we generalize that result in a simplified but time dependent setting, which is suitable for our discussion. 

The gauge freedom corresponding to $\F_i$ can conveniently be fixed by imposing
\be\label{gc1}
G_i=\del_j\cc_{ji}=0,
\ee
which obeys $\lp \F_i,G_j\rp\not=0$. This condition completely fixes the spatial diffeomorphisms generated by $k^i$. To preserve $G_i$ under the Hamiltonian flow, i.e. to have $dG_i/dt=\lp G_i,H\rp=0$, the Lagrange multiplier $n^i$ must be set to
\be\label{ni}
n^i=\fr{1}{2a^3}\fr{1}{\del^2}\del_iP_\z. 
\ee
To make the constrained phase space defined by the conditions $\F_i=G_i=0$ more obvious, one may use the decomposition
\bea
&&\cc_{ij}=\cc_{ij}^{TT}+2\del_{(i}\cc^T_{j)} +\del_i\del_j\cc-\fr13\d_{ij}\del^2\cc,\nn\\
&&\Pi_{ij}=\Pi_{ij}^{TT}+2\del_{(i}\Pi^T_{j)} +\del_i\del_j\Pi-\fr13\d_{ij}\del^2\Pi,\label{dec}
\eea
where $\cc_{ij}^{TT}$ and $\Pi_{ij}^{TT}$ are transverse-traceless tensors, and $\cc_i^T$ and $\Pi_i^T$ are transverse vector fields. The equations $\F_i=G_i=0$ imply 
\be\label{csur1}
\Pi^T_i=\cc_i^T=0,\hs{5}\cc=0,\hs{5}\Pi=-\fr{1}{4\del^2}P_\z.
\ee
The motion is now confined in this (partially) constrained phase space defined by \eq{csur1}. To obtain the reduced Hamiltonian ${\cal H}'$  generating the dynamics in this  subspace, one may use \eq{csur1} directly in  \eq{h} (see the appendix \ref{ap1}). Not surprisingly, the gauge condition \eq{gc1} decouples the dynamics of the tensor and the scalar sectors from each other giving ${\cal H}'={\cal H}_\cc+{\cal H}_S$,  where ${\cal H}_\cc$ and ${\cal H}_S$ are the respective Hamiltonians. We find that the Hamiltonian of the tensor modes becomes 
\be
{\cal H}_\cc=\fr{2}{a^3}\Pi_{ij}^{TT}\Pi_{ij}^{TT}+\fr18\,a\,\del_i\cc_{jk}^{TT}\del_i\cc_{jk}^{TT},
\ee
which corresponds to the standard action for the gravitational waves \eq{cceq}. 

On the other hand, the scalar sector involving the fields $\z$ and $\vf$ has the Hamiltonian 
\be
{\cal H}_S=-a\del_i\z\del_i\z-\fr{1}{2}\df\,\vf P_\z +\fr{1}{2a^3}P_\vf^2+\fr{1}{2}\,a\,\del_i\vf\del_i\vf+\fr12 a^3\overline{V}_{\f\f}\vf^2-\fr{3}{4}a^3\df^2\,\vf^2-a^3n\F\label{h2}
\ee
where the first class constraint is given by
\bea
\F=\,\ddot{\f}\,\vf-\fr{\df}{a^3} P_\vf-\fr{H}{a^3}P_\z-\fr{2}{a^2}\,\del^2\z.\label{cscalar}
\eea
As a consistency check, one may verify that $d\F/dt=\lp\F, H_S\rp+\del\F/\del t=0$. Note that for convenience we rescale  the constraint in \eq{h2} by $a^3$ as compared to \eq{h}. Time reparametrizations produced by the vector fields of the form $k^\m=(k,0)$ are generated by  $\Th=\int d^3 x\, a^3 \,k\,\F$. Gauge fixing can be done in various ways and as we discuss in the appendix \ref{ap2} it is possible to rederive the previously known results from \eq{h2} by imposing $\vf=0$ or $\z=0$ gauges. 

We would like to find a smooth gauge condition that  completely eliminates the inflaton fluctuation from the system, leaving $\z$ as the main scalar mode. Since \eq{cscalar} determines a linear combination of  $\vf$ and $P_\vf$, one should  impose a complementary condition that can be used to solve for $\vf$ and $P_\vf$.  We set
\be\label{cg}
G=c(t)\vf+d(t)P_\vf=0,
\ee
where $c(t)$ and $d(t)$ are two arbitrary functions. To have a unique solution for $\vf$ and $P_\vf$ from \eq{cscalar} and \eq{cg}, one needs
\be
\det \left[\begin{array}{cc}\ddot{\f} &-\fr{\df}{a^3}\\c&d\end{array}\right]\not=0
\ee
and without loss of any generality we impose
\be\label{cd}
\fr{\df}{a^3}\,c+\ddot{\f}\,d=1,
\ee
which normalizes the above determinant. During reheating when $\df=0$ one has $\ddot{\f}\not=0$, and vice versa. Consequently, it is possible to choose completely regular functions $c(t)$ and $d(t)$ obeying  \eq{cd} (see \eq{cd1} for an explicit example).  Since $\lp \F(t,x),G(t,y)\rp=\d^3(x-y)$, the function $G$ defines a viable one parameter family of smooth gauge conditions.\footnote{Note that given $d(t)$, the other function $c(t)$ can be determined from \eq{cd}. Below, we also reinterpret \eq{cg} in terms of the configuration space variables.} The constancy of $G$, i.e. $dG/dt=\lp G,H_S\rp+\del G/\del t=0$ determines the lapse uniquely as
\be\label{nn}
n=\left[d^2\overline{V}_{\f\f}-\fr32\df^2d^2+\fr{c^2}{a^6}+\fr{1}{a^3}(\dot{d}c-d\dot{c})-\fr{d^2}{a^2}\del^2\right]\left[\fr{H}{a^3}P_\z+\fr{2}{a^2}\del^2\z\right]-\fr{d\df}{2a^3}P_\z.
\ee
As we will see below, this expression greatly simplifies when expressed in the configuration space. 

It is straightforward to eliminate $\vf$ and $P_\vf$ by using \eq{cscalar} and \eq{cg}:
\bea
&&\vf=d\left[\fr{H}{a^3}P_\z+\fr{2}{a^2}\del^2\z\right],\nn\\
&&P_\vf=-c\left[\fr{H}{a^3}P_\z+\fr{2}{a^2}\del^2\z\right].\label{vps}
\eea
As we discuss in the appendix \ref{ap1}, the reduced Hamiltonian for the curvature perturbation\footnote{At the linearized level, $\z$ still determines the Ricci scalar of the induced metric $h_{ij}$ on the constant-$t$ hypersurfaces since $\del_i\cc_{ij}=0$.} $\z$ can be obtained by both plugging these solutions into the Hamiltonian \eq{h2} and adding the extra term emerging from a time dependent canonical transformation, which is derived in \eq{correction}. By comparing \eq{cscalar} and \eq{cg} with the corresponding equations \eq{a1} and \eq{a2}, one sees that  the functions $c(t)$ and $d(t)$ are identical and by comparing \eq{cscalar} with \eq{a1} one determines the function $K$ that appears in \eq{correction} as 
\be\label{k}
K= \left[\fr{H}{a^3}P_\z+\fr{2}{a^2}\del^2\z\right].
\ee
Using \eq{vps}, \eq{k} and \eq{correction}, the reduced Hamiltonian for $\z$ can be found as
\be\label{hz}
{\cal H}_\z=-a\del_i\z\del_i\z+\fr{a}{2}d^2\del_iK\del_iK+\fr{a\df}{H}d\,K\,\del^2\z+\fr12 \m a^3 \,K^2,
\ee
where the time dependent function $\m$ is given by
\be\label{mu}
\m=-\fr{\df}{H}d+\fr{1}{a^6}c^2+\overline{V}_{\f\f}\,d^2-\fr32\df^2 d^2+\fr{1}{a^3}\left[\dot{d}c-d\dot{c}\right].
\ee
Note that \eq{hz} contains spatial derivatives of the momentum variable $P_\z$ through \eq{k}, which is unconventional. 

Let $\zc$ be the curvature perturbation in the comoving gauge that has the action \eq{qac}. One can see that the following {\it time-dependent} linear canonical map 
\bea
&&\zc=\z-\fr{H^2d}{\df a^3}\left[P_\z+\fr{2a}{H}\del^2\z\right],\nn\\
&&P_{\zc}=P_\z+\fr{2a}{H}\del^2\z\label{canz}
\eea
transforms the Hamiltonian of $\zc$ to \eq{hz}, which proves the equivalence of the dynamics. 

The quadratic action corresponding to the Hamiltonian \eq{hz} can be obtained as
\be\label{ya}
S^{(2)}_\z=\int \fr12 M\left[\dz-\fr{\df d}{a^2}\del^2\z\right]^2-\fr{a\df^2}{2H^2}\del_i\z\del_i\z,
\ee
where
\be\label{mm}
M=\fr{a^3}{H^2} \left[\m-\fr{d^2}{a^2}\del^2\right]^{-1}.
\ee
In the Fourier space, $M$ must be evaluated by the substitution  $\del^2\to-k^2$, which fixes the notation in \eq{ya}.\footnote{Namely, the Green function must be understood to be sandwiched between the two terms in \eq{ya}.}  As we give an explicit example below, it is possible to choose $c$ and $d$ so that  $\m>0$ and $M$ becomes strictly positive. Although the action is nonlocal in the position space, it  involves only two time derivatives and the propagation for $\z$ is well defined once the initial conditions $\z(t_0,x)$ and $\dz(t_0,x)$ are given at some time $t_0$. Eq. \eq{ya} generalizes the action \eq{qac} of the comoving gauge. Indeed, setting $d=0$ in \eq{cd} gives $c=a^3/\df$ and \eq{ya} reduces to \eq{qac}. 

In obtaining the action \eq{ya} from \eq{hz}, a timelike integration by parts is applied. Due to the surface term thrown away in \eq{ya}, the original canonical momentum $P_\z$ is given by 
\be\label{lpz}
P_\z=\fr{\d S_\z^{(2)}}{\d \dz}-\fr{2a}{H}\del^2\z=M\left[\dz-\fr{\df d}{a^2}\del^2\z\right]-\fr{2a}{H}\del^2\z. 
\ee
This relation will be important in the next section. 

The equations of motion can be found  as
\be\label{yze}
\fr{\del}{\del t}\left[M\left(\dz-\fr{\df d}{a^2}\del^2\z\right)\right]= -\fr{d\df}{a^2}M\left[\del^2\dz-\fr{\df d}{a^2}\del^2\del^2\z\right]+\fr{a\df^2}{H^2}\del^2\z.
\ee
As long as $c$ and $d$ are chosen smoothly, $\m$ and $M$ become well behaved and one encounters no singularity in the $\z$ equation during reheating. Not surprisingly, the constant configuration 
\be
\z=\z_0
\ee
solves \eq{yze}. As pointed out  above, this is because of the fact that $\z$ is related to the global scaling of the metric function $a\to\l a$. Therefore, with the gauge condition \eq{cg} the curvature perturbation $\z$ becomes a well defined conserved variable during reheating. 

\section{\leftline{Implications}}\label{sec5}

From the expression for $P_\vf$ given in \eq{conm}, one may see that the gauge condition \eq{cg} can be written as $c\,\vf+a^3\,d\,[\dot{\vf}-\df \,n]=0$. Using this equation, it is easy to see that given an arbitrary set of perturbations $(\vf,n)$ one can apply the gauge transformation \eq{gt} with the parameter 
\be
k=-\fr{1}{a^3}\left[c\,\vf+a^3\,d\,\left(\dot{\vf}-\df \,n\right)\right]
\ee
so that \eq{cg} is satisfied (recall that the functions $c$ and $d$ are normalized to obey \eq{cd}). As a result, the gauge conditions introduced in the previous section can be written as 
\be\label{cdg}
\del_i\cc_{ij}=0,\hs{10}c\,\vf+a^3\,d\,\left[\dot{\vf}-\df \,n\right]=0.
\ee
We refer \eq{cdg} as the $cd$-gauge, which completely fixes the infinitesimal diffeomorphism invariance in the free theory. In this gauge, the basic variables are $(\cc_{ij},\z)$ and the other perturbations  $\vf$, $n$ and $n^i$ can be expressed in terms of $\z$. Indeed, using  the momentum $P_\z$ in \eq{lpz}, the lapse $n$ can be fixed from \eq{nn} as
\be\label{ncd}
n=\fr{\dz}{H}+\fr{\df d}{2a^3}M\left[\dz-\fr{\df d}{a^2}\del^2\z\right].
\ee
Similarly, \eq{ni} determines $n^i$ as
\be\label{nicd}
n^i=\fr{1}{2a^3}M\del_i\left[\del^{-2}\dz-\fr{\df d}{a^2}\z\right]-\fr{1}{Ha^2}\del_i\z
\ee
and \eq{vps} implies 
\be\label{fcd}
\vf=\fr{Hd}{a^3}M\left[\dz-\fr{\df d}{a^2}\del^2\z\right].
\ee
Note that only the derivatives of $\z$ appear in these expressions for $n$, $n^i$ and $\vf$. 

As emphasized above, the constant field $\z=\z_0$ solves the equations of motion. Neglecting the spatial  derivatives\footnote{It is possible to see that \eq{yze} admits a derivative expansion for $\m\not =0$.} in \eq{yze}, the second ``superhorizon" solution can also be obtained so that 
\be\label{01}
\z_k=c_1+c_2\int \fr{H^2}{a^3}\m\,dt\hs{10}k\simeq0.
\ee
As long as $c\propto a^3$ and $d$ is chosen independent of $a$, one can see from \eq{mu} that $\m$ becomes independent of $a$. Thus, the two solutions in \eq{01} can be referred as the constant and the decaying modes, and  unlike the solution given in \eq{0o} these are completely regular during reheating.

It is possible to transform the curvature perturbation $\z$ in the $cd$-gauge to the curvature perturbation in the comoving gauge, which has been denoted by $\zc$. Since the comoving gauge is defined by the condition $\vf=0$, one can apply a coordinate change to the perturbations in the $cd$-gauge that sets $\vf=0$. The corresponding gauge parameter $k$ in \eq{gt} can be found from \eq{fcd}, and applying the coordinate change gives\footnote{Eq. \eq{zcz1} can also be obtained from \eq{canz} and \eq{lpz}.}
\be\label{zcz1}
\zc=\z-\fr{H^2\,d}{\df\,a^3}\,M\,\left[\dz-\fr{\df d}{a^2}\del^2\z\right],
\ee
where $\z$ is the curvature perturbation in the $cd$-gauge. Similarly, starting with the comoving gauge with $\vf=0$ and the lapse given in \eq{ls}, one can apply a coordinate transformation to generate new fields $\vf$ and $n$ obeying \eq{cd}. Finding the corresponding parameter and changing the curvature perturbation yield
\be\label{zcz2}
\z=\zc+d\,\df\,\dot{\z}_c.
\ee
Note that \eq{zcz1} and \eq{zcz2} are inverse of each other provided that $\z$ and $\zc$ obey their respective equations of motion. We check that the lapse $n$ and the shift $n^i$, which are determined by the respective curvature perturbations, also transform accordingly  under these transformations. 

Since $\z$ is smooth, \eq{zcz1} shows that $\zc$ diverges like $1/\df$ as $\df\to0$, as it should be. On the other hand, defining a new variable $v=\df\, \zc$,  \eq{zcz2} becomes $\z=d\dot{v}+cv/a^3$, which shows that $\z$ is smooth since $v$ is well behaved. 

The power spectrum of $\zc$ can be obtained by quantizing the action \eq{qac}  on the inflationary background. In the slow-roll approximation with the Bunch-Davies vacuum, the constant superhorizon mode, which is denoted by $\z_k^{(0)}$ in \eq{0o}, becomes (see e.g. \cite{mfb}) 
\be\label{ps}
\zc{}_{k}\simeq \z_k^{(0)}=\fr{H(t_k)^2}{\df(t_k)}\fr{1}{\sqrt{2k^3}},
\ee
where $t_k$ is the time for horizon crossing: $k/a(t_k)=H(t_k)$ (the scalar power spectrum is defined by $|\z_k^{(0)}|^2$). Assuming further that the solutions in \eq{01} give the constant and the decaying modes respectively,\footnote{As noted above, this depends on the behavior of $c$ and $d$. For the functions given in \eq{cd1} below, this identification is true.} \eq{zcz2} shows that to a very good approximation the constant superhorizon modes are the same in the comoving and the $cd$-gauges:
\be
\z_k= \z_{c}{}_{k}=\z_k^{(0)},\hs{10}k\simeq0.
\ee
Namely, during inflation at superhorizon scales the initial $\z$ spectrum is identical to $\zc$ spectrum. As can be seen from \eq{zcz2}, changing the function $d$ only affects the decaying solution for the superhorizon modes. Consequently, our new variable $\z$ has the standard inflationary power spectrum and it safely propagates this initial scale-free spectrum beyond reheating. 

An important variable that has a direct physical meaning is the curvature perturbation in the longitudinal gauge, or the Newtonian gravitational potential\footnote{To be precise, the Newtonian potential arising from the linearization of general relativity around flat space is $-\np$.} $\np$, which determines the density perturbations.\footnote{The relation between $\np$ and $\d\rho$ can easily be obtained from the perturbed Einstein's equations.}  In the longitudinal gauge,  the metric takes the following form
\be
ds^2=-(1-2\np)dt^2+(1+2\np)a^2d\vec{x}^2
\ee
i.e. in terms of the perturbations introduced in \eq{cpert} one has 
\be
\z=\np,\hs{10}n=-\np,\hs{10}n^i=0.
\ee
It is possible to transform $\z$ or $\zc$ to $\np$ by finding the suitable parameter $k$ that sets $n^i=0$ in the transformation \eq{gt}, which gives
\be\label{npz}
\np=\left\{\begin{array}{ll}\fr{H}{2a}M\,\left[\del^{-2}\dz-\fr{\df d}{a^2}\z\right]\hs{7}cd-\textrm{gauge}\\  \fr{\df^2a^2}{2H}\del^{-2}\dot{\z}_c\hs{23}\textrm{comoving gauge}.\end{array}\right.
\ee
Eq.  \eq{npz} shows that  the superhorizon mode $\np_k$ is fixed by the leading order derivative correction of the constant superhorizon curvature perturbation  and an easy calculation gives
\be\label{npf}
\np_k\simeq\left[\fr{H}{2a}\int^t a(t')\fr{\df(t')^2}{H(t')^2}dt'\right]\z_k^{(0)},\hs{7}k\simeq0.
\ee
Note that the Newtonian potential $\np$ is well defined at all epochs.  

Let us finally consider an explicit example for the $cd$-gauge that would illustrate some of the points above. We first specify the background evolution for a reasonably large set of generic models. During inflation, the background dynamics  is controlled by the standard slow-roll parameters defined by
\be
\e=\fr12\fr{V_\f^2}{V^2},\hs{10}\eta=\fr{V_{\f\f}}{V}.
\ee
We take a massive inflaton with mass $m$, therefore the potential in reheating can be taken as $V\simeq \fr12 m^2\phi^2$. Assuming further that  $m\gg H$, which is generically true in many models, the scalar field equation in \eq{be} can be solved approximately as  
\be\label{b1}
\phi(t)\simeq\Phi \sin(mt),
\ee
where the slowly changing amplitude obeys $\dot{\Phi}+3H\Phi/2\simeq0$. In that case the background equations yield the following approximate solution 
\be \label{fh}
a\simeq\left(\fr{t}{t_0}\right)^{2/3},\hs{5}\Phi\simeq\fr{t_0}{t}\F_0,
\ee
which is valid during reheating.

A convenient choice for the functions $c$ and $d$  is 
\be\label{cd1}
c=\fr{m^2\,a^3\,\df }{m^2\df^2+\ddot{\f}^2},\hs{10}d=\fr{\ddot{\f}}{m^2\df^2+\ddot{\f}^2},
\ee
which satisfies the normalization condition \eq{cd}. During the exponential expansion, these functions are determined by the slow-roll parameters as
\be
c\simeq-\fr{a^3}{H\sqrt{2\e}},\hs{10}d\simeq\fr{\eta-\e}{m^2\sqrt{2\e}}.
\ee
On the other hand,  \eq{fh} can be used to obtain the form of the functions in reheating
\be
c\simeq\fr{a^3\cos(mt)}{m\Phi},\hs{10}d\simeq-\fr{\sin(mt)}{m^2\Phi},
\ee
which are completely well behaved.  From \eq{cd1},  the second superhorizon solution given in \eq{01} can be seen to be decaying like $1/a^3$ during inflation. 

Another suitable  choice is (recall that we are using the Planck units  $8\pi G=1$) 
\be\label{cd2}
c=\fr{a^3\,\df }{\df^2+\ddot{\f}^2},\hs{10}d=\fr{\ddot{\f}}{\df^2+\ddot{\f}^2}\,\, .
\ee
Using \eq{b1} and \eq{cd2} in \eq{mu}, we numerically check for a various phenomenologically interesting set of parameters that one has a strictly positive function $\m>0$ during reheating. 

In this model, one can use \eq{npf} to relate the Newtonian gravitational potential to the initial inflationary power spectrum. Using \eq{b1} in \eq{npf}  and remembering that we have assumed $m\gg H$, the highly oscillatory integral can be approximated by treating the slowly changing functions as constants that gives
\be\label{sr}
\np_k\simeq\fr35\z_k^{(0)},
\ee
which  is the standard relation between $\np_k$ and $\z_k^{(0)}$ in a matter dominated universe. This calculation proves that there cannot be any amplification of  the superhorizon density perturbations $\d\rho_k$ if the inflaton is massive since $\np_k$ is constant.  This is a highly nontrivial result that has been shown in \cite{rme1} by carefully analyzing the equations of motion for perturbations. We see that the same conclusion can straightforwardly be reached by using the conserved variable  $\z$. 

\section{\leftline{Conclusions}}

The conservation of the superhorizon curvature perturbation is crucial in directly relating the late time cosmological observables like the CMB temperature perturbations to the quantum mechanical vacuum fluctuations in the very early universe during inflation. The general belief is that due to the conservation law the details of the cosmic evolution after inflation does not matter for the late time properties of the cosmological perturbations. However, it is not much appreciated in the literature that the standard curvature perturbation in the comoving gauge becomes singular during reheating if the inflaton oscillates about the minimum of its potential. 

In this paper, we try to solve this problem by introducing smooth alternative gauges, which are both regular in reheating and still eliminate the inflaton perturbation from the dynamics. As we have shown, this can be achieved in the Hamiltonian formalism where the coordinates and momenta are treated as independent variables. Although it obeys an unconventional equation of motion, the new curvature perturbation becomes a smooth variable in reheating whose superhorizon mode is rigorously conserved.  

It turns out that the new gauge can be chosen in such a way that the constant superhorizon mode in inflation is equal to the corresponding mode in the comoving gauge. In other words, only the sub-leading decaying pieces differ between gauges. Therefore, the new curvature perturbation has the standard inflationary power spectrum and using its conservation law it is possible to propagate the initial scale free spectrum beyond reheating. For instance, an important relation is \eq{sr} that relates the Newtonian potential and hence the density perturbation to the initial inflationary spectrum. In early works (see e.g. \cite{rme2}), \eq{sr} has been obtained from the conservation of the (singular) curvature perturbation by approximating the coherent inflaton oscillations by a matter dominated universe on the average. Here, we obtain this result in a rigorous way without referring  to the singular variable. 

In this work, we have focused on the reheating era and assumed that the perturbations during inflation can be treated in the standard way. However, there has been some concerns about the behavior of the cosmological perturbations in the limit $\df\to0$ as the nearly exponential expansion approaches to the exact de Sitter space, since the standard power spectrum \eq{ps} diverges (see e.g. \cite{gr}). For the moment, our findings cannot be directly applied to analyze this issue since we assume that $\ddot{\f}\not=0$ when $\df=0$ and vice versa. Nevertheless, it would be interesting to work out the necessary modifications in an attempt to clarify $\df\to0$ limit of inflation. 

\appendix

\section{Gauge fixing and  phase space reduction in the constrained Hamiltonian systems}\label{ap1}

In this appendix, we consider a classical mechanical system whose dynamics is determined by a Hamiltonian and a first class constraint. We assume that the constraint is at least linear in a coordinate and the conjugate momentum, and otherwise arbitrary. We study the evolution after gauge fixing in two equivalent ways; either by using the Dirac's method of the constrained systems \cite{dirac} or by reducing the dynamics on the constrained phase space. Our analysis is a simplified version of \cite{cs}, however we generalize that discussion by taking both the Hamiltonian and the constraint to have explicit time dependencies, which is suitable for cosmological applications. We show that the two methods are equivalent, i.e. they yield identical evolutions. 

Let us denote the Hamiltonian as $H(q,p,q_i,p_i;t)$. The first class constraint is assumed to have the form 
\be\label{a1}
\F=a(t)q+b(t)p-K(q_i,p_i;t),
\ee
where $K$ is an arbitrary function and $a(t)$ and $b(t)$ denote time dependent external parameters.  Being a first class constraint, $\F$ must obey $d\F/dt=\lp \F,H\rp+\del\F/\del t=0$. The total Hamiltonian that generates  the most general motion  can be written as 
\be\label{aht}
H_T=H+\l\F,
\ee
where $\l$ is a Lagrange multiplier that is completely arbitrary at the moment. 

Due to the existence of the constraint, the evolution of the system is actually confined in a codimension one sub-manifold defined by $\F=0$. Besides,  $\F$ generates a gauge transformation and one may fix that freedom to reduce the motion in a codimension two manifold, which can be viewed as a reduced phase space. The gauge freedom can be fixed by introducing a function $G$ that obeys $\lp G,\F\rp\not=0$. In that case the codimension two surface in which the motion is confined is given by $\F=G=0$. For our gauge condition we take 
\be\label{a2}
G= c(t)q+d(t)p=0
\ee
and without loss of  any generality impose $a(t)d(t)-b(t)c(t)=1$ so that $\lp \F,G \rp=1$. As mentioned above,  to study the dynamics of the system, one may either follow Dirac, which amounts to determine  the Lagrange multiplier $\l$ from the condition $dG/dt=0$, or solve for the conditions \eq{a1} and \eq{a2} to find the reduced codimension two phase space and the corresponding reduced Hamiltonian. 

In the first case, $dG/dt=\lp G,H_T\rp+\del G/\del t=0$ fixes  $\l$ as  
\be\label{al}
\l=-\fr{1}{\lp G,\F\rp}\left[\lp G, H\rp +\fr{\del G}{\del t}\right]=-c(t)\fr{\del H}{\del p}+d(t)\fr{\del H}{\del q}-\dot{c}(t)q-\dot{d}(t)p,
\ee
and the equations of motion become
\be
\dot{q}_i=\fr{\del H}{\del p_i}+\l\fr{\del K}{\del p_i},\hs{10}\dot{p}_i=-\fr{\del H}{\del q_i}-\l\fr{\del K}{\del q_i}.\label{ade}
\ee
The evolutions of $q$ and $p$ are determined by the conditions $\F=G=0$, which read $q=d(t)K(q_i,p_i;t)$ and $p=-c(t)K(q_i,p_i;t)$.

Let us now try to obtain a gauge fixed Hamiltonian $\tilde{H}(q_i,p_i;t)$ that describes the dynamics of the ``true physical degrees of freedom"  $q_i$ and $p_i$. Of course, the flow generated by $\tilde{H}(q_i,p_i;t)$ must be identical to \eq{ade}. Since $\F=G=0$ implies $q=d(t)K(q_i,p_i;t)$ and $p=-c(t)K(q_i,p_i;t)$, one may first  try a direct substitution in the Hamiltonian \eq{aht}, which gives $H(d(t)K(q_i,p_i;t),-c(t)K(q_i,p_i;t),q_i,p_i;t)$. But  it is easy to see that the equations generated by this Hamiltonian misses the terms involving $\dot{c}(t)$ and $\dot{d}(t)$ in \eq{ade} through \eq{al}. Here, the explicit time dependence of the gauge condition causes a problem. 

To resolve this issue, one may first apply the following canonical transformation 
\bea
Q=a(t)q+b(t)p,\nn\\
P=c(t)q+d(t)p,\label{act}
\eea
which can be generated by the function 
\be
F_1(q,Q,t)=-\fr{a(t)}{2b(t)}q^2-\fr{d(t)}{2b(t)}Q^2+\fr{1}{b(t)}qQ,
\ee
where one has $p=\del F_1/\del q$ and $P=-\del F_1/\del Q$. Under this transformation, the Hamiltonian must also be shifted as
\be\label{ah2}
H'(Q,P,q_i,p_i;t)=H(q(Q,P),p(Q,P),q_i,p_i)+\fr{\del F_1}{\del t},
\ee
where the substitutions in $F_1$ must be done after the partial time derivative is evaluated. In these new canonical variables $(Q,P,q_i,p_i)$, the constraint $\F$ and the gauge fixing condition $G$ become
\bea
&&Q=K(q_i,p_i;t),\nn\\
&&P=0.\label{act2}
\eea
In these equations the new canonical coordinates, which are going to be eliminated, are not multiplied by time dependent factors  and one may  obtain the reduced Hamiltonian on the constrained surface by simply using \eq{act2} in \eq{ah2}, which yields 
\be\label{correction}
\tilde{H}(q_i,p_i;t)=H\left(d(t)K(q_i,p_i;t),-c(t)K(q_i,p_i;t),q_i,p_i;t\right)+\fr12\left[\dot{d}(t)c(t)-\dot{c}(t)d(t)\right]K(q_i,p_i;t)^2.
\ee
It is easy  to check that the equations generated by this Hamiltonian are identical to \eq{ade}. Eq. \eq{correction} is the basic result of this appendix, which is used in the main text above. 

\section{Rederiving the equations in the previously known gauges}\label{ap2}

In this section, we rederive the well known equations corresponding to the standard gauges $\vf=0$ and $\z=0$ from \eq{h2}. Let us start by imposing $\vf=0$ which corresponds to the comoving gauge discussed in section \ref{sec2}. To eliminate the inflaton field completely, one may use the constraint to solve for $P_\vf$ as
\be
P_\vf=-\fr{H}{\df}P_\z-\fr{2a}{\df}\del^2\z.
\ee
As we discuss in the appendix \ref{ap1},  the reduced Hamiltonian can be obtained by setting $\vf=0$ and using the solution for $P_\vf$ in \eq{h2} that yield
\be
{\cal H}_\z=-a\del_i\z\del_i\z+\fr{1}{2a^3\df^2}\left[HP_\z+2a\del^2\z\right]^2.
\ee
It is straightforward to show that the corresponding Lagrangian is equal to \eq{qac}. Moreover, one can check that the solutions for the Lapse $n$, which can be obtained from $\lp \vf,H_S\rp=0$, and the shift $n^i$ given in \eq{ni} are identical to \eq{ls}. 

Similarly, it is also possible to impose $\z=0$ and use \eq{cscalar} to solve for $P_\z$ as
\be
P_\z=\fr{a^3\ddot{\f}}{H}\vf-\fr{\df}{H}P_\vf.
\ee
Setting $\z=0$ and using this solution in \eq{h2} give
\be
{\cal H}_\vf=\fr{1}{2a^3}P_\vf^2+\fr{a}{2}\del_i\vf\del_i\vf+\fr{\df^2}{2H}P_\vf\vf+\fr{a^3}{2}\left[\overline{V}_{\f\f}-\fr32\df^2-\fr{\df\ddot{\f}}{H}\right]\vf^2.
\ee
After a Legendre transformation, one then obtains
\be
S^{(2)}=\int\fr{a^3}{2}\left[\dot{\vf}^2-\fr{1}{a^2}\del_i\vf\del_i\vf-\fr{\df^2}{H}\vf\dot{\vf} +\left(\fr{\df^4}{4H^2}+\fr32\df^2-\overline{V}_{\f\f}+\fr{\df\ddot{\f}}{H}\right)\vf^2\right],
\ee 
which is the standard action for the inflaton fluctuation, see e.g. \cite{mal}. It can be further checked that the shift $n^i$ and the lapse $n$, which are obtained from \eq{ni} and from the condition $\lp \z,H_S\rp=0$, also match the usual expressions given in \cite{mal}. Consequently, we show that  it is possible to obtain the well known equations for the  cosmological perturbations in $\vf=0$ and $\z=0$ gauges from \eq{h2}. 

\acknowledgments

We thank Richard Woodard for a helpful e-mail correspondence and for pointing out the reference \cite{gr}.

\end{document}